\DocumentMetadata{
  lang        = en,
  pdfstandard = ua-2,
  pdfstandard = a-4,
}

\documentclass[11pt,a4paper,twoside=off, DIV=11]{scrartcl}

\usepackage[T1]{fontenc}
\usepackage[utf8]{inputenx}
\usepackage{lmodern}
\usepackage{scrlayer-scrpage}
\usepackage{authblk}
\usepackage{lettrine}
\usepackage{graphicx}
\usepackage{amsmath}
\usepackage{amssymb}
\usepackage{amsfonts}
\usepackage{booktabs}
\usepackage[english]{babel}

\addtokomafont{disposition}{\rmfamily}

\setkomafont{pagehead}{\footnotesize}

\ihead{IEEE IWCMC}
\ohead{May 2025}

\begin{document}

\title{A Comparative Study of Recent Advances in Internet of Intrusion Detection Things}

\author[1]{Marianna Rezk}
\author[2]{Hassan Harb}
\author[1]{Ismail Bennis}
\author[1]{S\'{e}bastien Bindel}
\author[1]{Abdelhafid Abouaissa}
\affil[1]{IRIMAS \\ University of Upper-Alsace, Mulhouse, France \newline \textit{marianna.rezk@uha.fr, ismail.bennis@uha.fr, sebastien.bindel@uha.fr, abdelhafid.abouaissa@uha.fr}}
\affil[2]{College of Engineering and Technology, American University of the Middle East, Kuwait \textit{hassan.harb@aum.edu.kw}}
\date{}
	
\maketitle	
\begin{abstract} 
The Internet of Things (IoT) has revolutionized the way devices communicate and interact with each other, but it has also created new challenges in terms of security. In this context, intrusion detection has become a crucial mechanism to ensure the safety of IoT systems. To address this issue, a comprehensive comparative study of advanced techniques and types of IoT intrusion detection systems (IDS) has been conducted. The study delves into various architectures, classifications, and evaluation methodologies of IoT IDS. This paper provides a valuable resource for researchers and practitioners interested in IoT security and intrusion detection. 
\end{abstract}       
	
\section{Introduction}
\lettrine{I}{n} today's interconnected digital landscape, securing information systems and networks is a critical task. To address these issues, Intrusion Detection Systems (IDSs) have emerged in order to detect and respond to unauthorized or malicious activities in host and network environments, contributing to keep their integrity~\cite{NNN001}. It performs monitoring network traffic and system behavior to identify any anomalies or patterns indicative of possible security breaches. As a result, it plays a significant role in incident response, threat analysis, and post-incident forensics. In the context of IoT, wherein all nodes are connected to the Internet and may send sensitive data, security challenges arise \cite{NNN002}. In addition, nodes are designed to minimize the energy consumption and without taking into account any security issues. To address these concerns, researchers attempt to integrate IDS into IoT networks. For example, some researchers are developing lightweight IDS that can run on IoT devices themselves, allowing them to detect and respond to threats in real-time. Others are exploring ways to use machine learning to improve the accuracy of IDS, allowing them to detect more sophisticated attacks. Overall, the development of IDS for IoT networks is an important area of research, as it could help to mitigate the security risks associated with the increasing connectivity of devices. While there is still much work to be done in this area, the potential benefits of such systems make them an exciting area of future work.

In this paper is detailed a comparative study of the detection performance of IDS using the NSL-KDD dataset\cite{kdd-nsl}. This one is widely recognized in the field of intrusion detection research for its diversity in network traffic scenarios and comprehensive coverage of attack types. The detection accuracy of considered IDS is assessed through basic and composite metrics as described in \cite{10.1145/2808691}. At the end, a final comparison is computed with the Friedman test. This statistical approach enabled us to identify significant differences in performance across multiple metrics, providing objective insights into the strengths and weaknesses of each technique.

The remaining sections of the paper are structured as follows. Section \ref{sec:AC} presents the main architecture of IDSs in the IoT context and the classification of existing methods in this domain and what are the main applications of IDS in IoT. Then, in section \ref{sec:BS}, we introduce the comparative study using efficient techniques to perform the comparison. In section \ref{sec:ECD}, we evaluate the results of the comparison between the selected approaches. Finally, in section \ref{sec:CFW}, we conclude our work.

\section{Architecture, Classification, Applications}\label{sec:AC}
\subsection{IoIDT: System Architecture}
Mostly, the IDS architecture within IoT networks is characterized by three interlinked layers: perception, network, and decision. This distinctive framework, comprising these three layers, forms a robust defense mechanism, specifically designed to address the unique security challenges associated with Internet of Intrusion Detection Things (IoIDT).

\paragraph{Perception Layer:}it deals with the initial data acquisition and preprocessing by collecting and analyzing raw data generated from IoT devices. Sensors within this layer monitor the network traffic, device logs, and other relevant data to identify patterns, anomalies, or potential security incidents.

\paragraph{Network Layer:}it builds upon the insights gained in the perception layer and focuses on analyzing the overall network traffic and communication patterns. It aims to identify any abnormal activities or deviations from established baselines. This layer may use techniques such as signature-based detection and anomaly-based detection to flag potential security threats. 

\paragraph{Decision Layer:}it is responsible for making decisions based on the analysis performed in the Perception and Network Layers. It determines whether a detected activity is a legitimate action or a potential security threat. Decision-making in this layer may involve comparing observed patterns against predefined rules, policies, or machine learning models trained on historical data. Depending on the severity and nature of the identified threat, the Decision Layer may trigger alerts, initiate automatic responses (such as blocking malicious traffic), or notify human operators for further investigation and intervention.

\begin{figure}[h]
  \centering
  \includegraphics[width=0.8\textwidth]{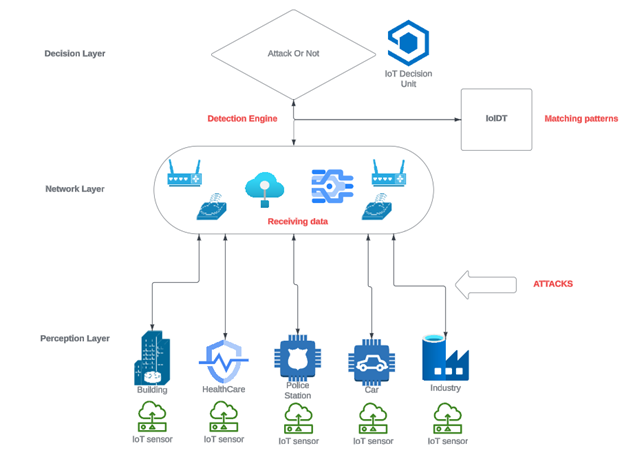}
  \caption{Architecture}\label{fig:architecture}
\end{figure}

The integration of these three layers creates a holistic approach to intrusion detection, allowing the system to adapt to the dynamic nature of IoT environments (Figure \ref{fig:architecture}). 

\subsection{Classification}
A common classification of IoT-enabled IDS split systems into two groups: signature-based and anomaly-based \cite{Khraisat2019}. On one hand, the signature-based IDS are based on matching patterns technique to detect attacks by comparing an attack with the signature of a previous ones stored in an internal database. This technique is effective to detect known threats but not for new ones, such as zero-days and variant of old threats. On the other hand, the anomaly-based IDSs establish a baseline of normal behavior and flag any deviation from it as potentially indicative of an intrusion. One can distinguish between three types of anomaly-based IDSs.

\paragraph{Behavioral Anomaly Detection:} by analyzing the typical behavior of devices such as sensors, actuators, and connected appliances, deviations from established patterns can be detected and potential security breaches can be identified. Baseline behavior models are established through continuous monitoring of device interactions and data transmissions, allowing for the detection of anomalies such as sudden spikes in sensor data or unusual device access patterns. These anomalies can trigger alerts, alerting users to potential unauthorized access or malicious activities. 

\paragraph{Statistical Anomaly Detection:} it is particularly effective in IoT environments where a vast amount of data is generated by interconnected devices. This method involves analyzing the statistical characteristics of device interactions, sensor readings, or network traffic to pinpoint any anomalies or outliers. By creating a statistical profile of normal behavior based on historical data, statistical anomaly detection can identify deviations that may indicate security breaches or abnormal activity. 

\paragraph{Machine Learning-based Anomaly Detection:} using machine learning-based anomaly detection algorithms can help to identify abnormal activities and recognize potential attacks in IoT data. These algorithms can analyze patterns and relationships within IoT data streams using supervised or unsupervised learning techniques. They can detect previously unseen threats or anomalies, by continuously refining their models with real-time data.

\subsection{IDS Application's in IoT}\label{AI}
\begin{figure}[h]
  \centering
  \includegraphics[width=0.5\textwidth]{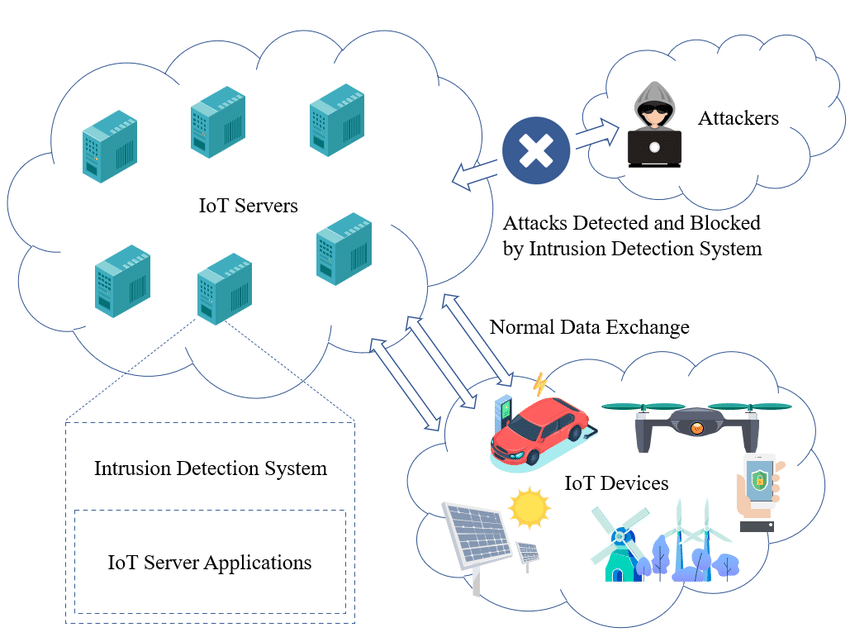}
  \caption{Applications of IDS in IoT \cite{a11}}\label{fig:applications}
\end{figure}

The effective management of security challenges presented by the Internet of Things (IoT) landscape is increasingly becoming a critical concern, necessitating the deployment of reliable and robust security measures such as Intrusion Detection Systems (IDS). The production of interconnected IoT devices across diverse domains presented in Figure \ref{fig:applications} like healthcare, transportation, manufacturing, and smart cities has heightened the need for strong security mechanisms to prevent potential data breaches, system abuse, or unauthorized access. Traditional security mechanisms often fall short of adequately protecting IoT environments because of their scale, heterogeneity, and dynamic nature. As such, IDS tailored explicitly for IoT have been designed to address these specific challenges.

We will explore the practical applications of Intrusion Detection Systems (IDS) in Internet of Things (IoT) environments, with a particular highlight on anomaly detection and signature-based detection methods.

\subsubsection{Anomaly Detection Application :} 
\begin{figure}[h]
  \centering
  \includegraphics[scale=0.3]{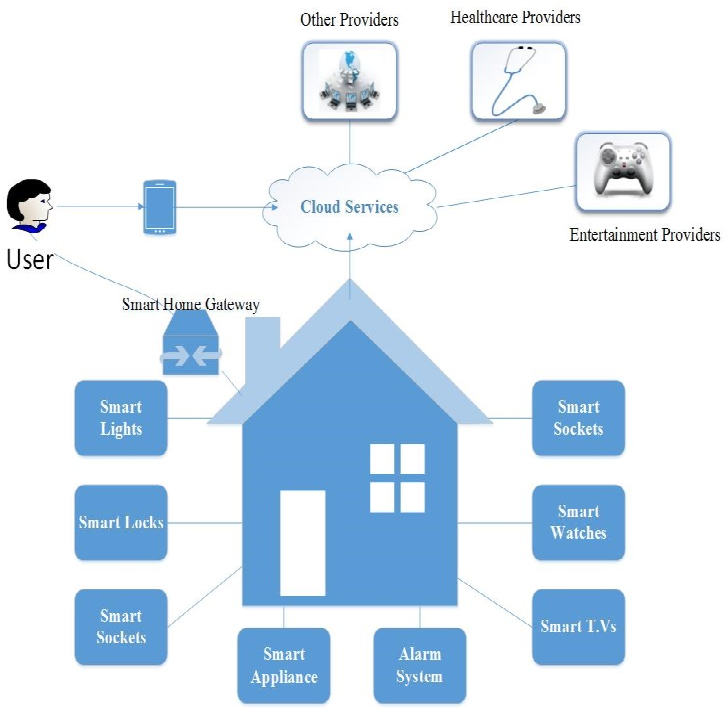}
  \caption{Smart Home Security \cite{a12}}\label{fig:smarthome}
\end{figure}

the example scenario that we are discussing is related to smart home security.
Smart Home Security systems with IoT devices like cameras, door sensors, and smart appliances require anomaly detection IDS to enhance security. The IDS closely monitors network traffic and device behavior in the smart home ecosystem for potential threats. The IDS conducts a thorough analysis of network traffic patterns and device behavior, taking note of customary actions such as periodic image uploads by cameras and alert triggers by door sensors. Any deviations from established norms, like irregular data transfer rates or unanticipated access patterns (e.g., excessive data uploads by a camera outside of regular intervals), prompt the IDS to flag suspicious activities. Upon detecting anomalies, the IDS generates alerts, which are usually communicated via mobile applications or email notifications to homeowners or security personnel. Depending on the severity of the situation, automated response measures may be activated, like isolating the affected device from the network to prevent potential threats while further investigation is conducted. The authors of \cite{a12} explain in Figure \ref{fig:smarthome} a central smart home gateway serves as a communication bridge between disparate smart devices within a residence, enabling both cloud-based control and automated functionalities.

\paragraph{Signature-Based Detection Application :} the example scenario that we are discussing is related to Industrial IoT (IIoT) Manufacturing System. In a manufacturing facility utilizing IIoT devices such as PLCs and industrial sensors, it is essential to have reliable IDS in place to protect important industrial operations from cyber threats.

\begin{figure}[h]
  \centering
  \includegraphics[width=0.5\textwidth]{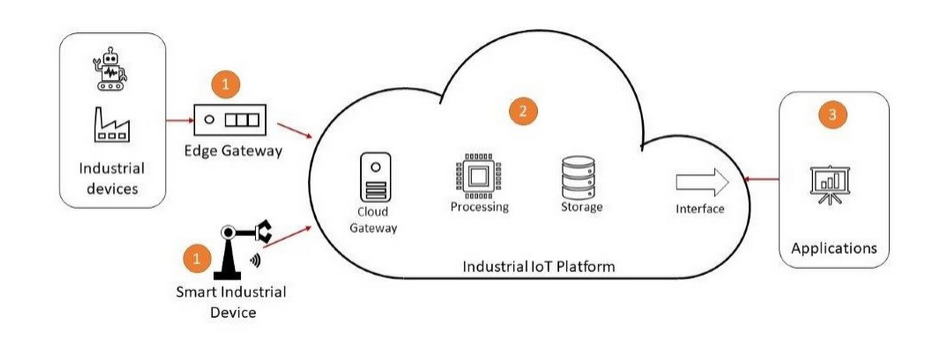}
  \caption{Industrial IoT Platform}
  \label{fig:industrial}
\end{figure}

Th eIDS comes equipped with a comprehensive database of attack signatures that are specifically tailored to industrial control systems (ICS) and IIoT devices. As data travels between PLCs, sensors, and supervisory control systems, the IDS performs real-time packet inspection by cross-referencing incoming data packets with its database of attack signatures. If the IDS detects a match with a known attack signature (such as a particular command sequence associated with a known PLC malware attack), it initiates an alert, indicating a potential intrusion attempt. This triggers predefined response actions, which may include blocking suspicious communication, logging detailed forensic data for further analysis, and notifying relevant staff or security teams. An IDS protects an Industrial IoT (IIoT) system in the Figure \ref{fig:industrial}. IIoT refers to the interlinking of industrial devices and machines over computer networks. These machines gather and share data, making it possible to monitor and control them remotely. An IDS is a security mechanism that checks a network for malicious activities or security breaches. It can identify and notify suspicious activities such as unauthorized access attempts, denial-of-service attacks, or data manipulation attempts.

These examples highlight the crucial role played by IDS utilizing both anomaly detection and signature-based detection methods in safeguarding diverse IoT systems against emerging cyber threats. The practical significance of IDS implementation in mitigating the unique security challenges faced by IoT deployments across various industry domains is evident.

\section{Intrusion Detection System: A Comparative Study}\label{sec:BS}
This section aims to identify and select the most outstanding research papers for in-depth examination and comparison using a variety of performance metrics and criteria. Our goal is to evaluate the effectiveness and robustness of different methods or approaches through quantitative measures, including accuracy, recall, precision, F1-score, False Positive Rate (FPR), and specificity. By analyzing each of these metrics, we intend to clarify the objectives of each research paper, facilitating later comparisons.

\subsection{Used Metrics: Explanation}
In the context of binary classification, the following terms are commonly used to evaluate the performance of any classifier.
  
\paragraph{Accuracy:}in the realm of IDS, accuracy stands as a pivotal metric, serving as a fundamental indicator of the system's proficiency in distinguishing between normal and malicious network activities. Accuracy, in the context of an IDS, encapsulates the system's ability to correctly identify and classify both intrusions and benign events. This metric is computed by summing up the true positives (instances where the system correctly identifies intrusions), and true negatives (instances where the system accurately recognizes normal activities), and then dividing this sum by the total number of instances.

\paragraph{Recall:}recall stands as a critical metric revealing the system's ability to identify and capture all instances of intrusions, minimizing the occurrence of false negatives. Also known as sensitivity or true positive rate, recall provides a focused evaluation of the system's capacity to detect and correctly label malicious activities amid the countless number of network events. The calculation of recall involves dividing the number of true positives (instances where the system correctly identifies intrusions) by the sum of true positives and false negatives (instances where the system fails to recognize actual intrusions). 

\paragraph{Precision:}it serves as a crucial indicator of the system's accuracy in flagging potential intrusions while minimizing false positives. Also known as positive predictive value, precision focuses on the proportion of correctly identified intrusions among all instances flagged as positive by the system. The precision metric is calculated by dividing the number of true positives (instances correctly identified as intrusions) by the sum of true positives and false positives (instances where the system incorrectly labels normal activities as intrusions). 

\paragraph{F1-Score:} it provides a balanced assessment of a system's performance by considering both precision and recall. This metric is particularly valuable in scenarios where achieving stability between minimizing false positives and false negatives is of utmost importance.

\paragraph{False Positive Rate:}it is shedding light on the system's propensity to generate false alarms by incorrectly identifying normal network activities as potential intrusions. Also known as the false alarm rate or Type I error rate, the FPR is a measure of the proportion of benign instances that are mistakenly flagged as intrusions. Mathematically, the False Positive Rate is calculated by dividing the number of false positives (instances where the system incorrectly identifies normal activities as intrusions) by the sum of false positives and true negatives (instances where the system accurately recognizes normal activities).

\paragraph{Specificity:}it is offering a targeted assessment of the system's ability to accurately identify and classify normal network activities while minimizing false positives. Often referred to as the true negative rate, specificity calculates the proportion of actual normal instances correctly identified by the system. The specificity metric is calculated by dividing the number of true negatives (instances where the system accurately recognizes normal activities) by the sum of true negatives and false positives (instances where the system incorrectly labels normal activities as intrusions).
  
\subsection{Selected Articles: Summary}
The selected articles in our study were carefully chosen for their relevance and contributions to the evolving domain of intrusion detection systems (IDS), particularly in addressing the complexities of IoT networks and modern cybersecurity challenges. These works encompass an ensemble of advanced methodologies, including bio-inspired optimization, deep learning architectures, feature selection strategies, and ensemble learning techniques, while tackling critical issues such as data imbalance and resource constraints in IoT environments. By integrating state-of-the-art approaches and innovative solutions, these studies provide a robust foundation for advancing IDS research and offer valuable insights into enhancing detection accuracy, scalability, and efficiency in real-world applications.
  
Nadir et al. \cite{a1} introduced the Firefly Optimization (FFO) technique for intrusion detection, preceding the application of a machine learning classifier for categorization. The study's experimentation utilized the Knowledge Discovery Dataset (KDD-CUP 99), encompassing 30,787 instances in the training phase and 29,954 instances in the testing phase. This dataset comprises 41 attributes, including a tag denoting whether an assault was perpetrated by a typical attacker or a specific type of attacker. The Firefly Optimization (FFO) algorithm, employed as part of the intrusion detection system, leverages the bio-inspired behavior of fireflies to enhance the search for optimal solutions in the detection process. Subsequently, the Probabilistic Neural Network was employed to categorize the database into anomalous and normal scenarios. This network calculates the likelihood that a given sample belongs to a recognized category and is structured with four layers: Input layer, Pattern layer, Summation layer, and Output layer.

A Deep Learning-Based Intrusion Detection System (IDS) is a security solution that leverages deep neural network architectures to automatically and intelligently detect and respond to unauthorized or malicious activities within a computer network. The authors of \cite{a2} are focusing on enhancing the performance of DNN-based IDS by proposing a novel feature selection technique that selects features based on statistical interpretation of features for the underlying intrusion detection datasets, such as NSL-KDD, UNSW\_NB-15, and CIC-IDS-2017, using Standard Deviation and Difference of Mean and Median. So here, as basic idea, DNN applied for learning and classification process using reduced feature subset. This integration of deep neural networks enhances the IDS's capability to adapt and evolve, providing a robust defense mechanism against a wide range of cyber threats. 

Data imbalance is a common challenge in the context of Intrusion Detection Systems (IDS). Imbalance occurs when the distribution of classes (normal and malicious activities) in the training dataset is uneven, with one class significantly outnumbering the other. In the case of IDS, the normal class (normal network behavior) tends to be much more prevalent than the malicious class (intrusions or attacks). To overcome this issue, the authors of \cite{a3} proposed the specialized loss function, called focal loss, that automatically down-weighs easy examples and focuses on the hard negatives by facilitating dynamically scaled-gradient updates for training effective ML models. The implementation of this model has been done using two deep learning neural network architectures: (i) Feed Forward Neural Networks (FNNs) and (ii) Convolutional Neural Networks (CNNs). The experiments are done using two public datasets: Bot-Iot and WUSTL-EHMS-2020.
 
In \cite{a4}, the authors propose a federated sampling and lightweight intrusion-detection system for IoT networks that use K-means for sampling network traffic and identifying anomalies in a semi-supervised way. The system is designed to preserve data privacy by performing local clustering on each device and sharing only summary statistics with a central aggregator. The proposed system is particularly suitable for resource-constrained IoT devices such as sensors with limited computational and storage capabilities.

Meanwhile, \cite{a5} propose stacked ensemble learning model for network intrusion detection by using gradient boosting machine (GBM) and random forest (RF) algorithms. The proposed method integrates the features of both GBM and RF classifiers. Random Forest is an ensemble machine learning algorithm that constructs a multitude of decision trees during training and outputs the mode of the classes (classification) or the mean prediction (regression) of the individual trees. Gradient Boosting is a machine learning technique that builds a predictive model in the form of an ensemble of weak learners, typically decision trees, sequentially. It aims to correct errors made by the previous models in the ensemble by assigning higher importance to the misclassified instances, ultimately creating a strong predictive model with improved accuracy and performance. Gradient boosting can do regression, classification and ranking.

\section{Evaluations, Comparison, and Discussions}\label{sec:ECD}
After choosing the five articles, and explaining each one in the previous section, now we want to analyze each one by taking into consideration their architectures, methodologies, advantages, and their limitations. For this purpose, we choose six different metrics explained in a previous section, where some of them are already defined in the desired article, and the rest is being executed and calculated. To implement the methodologies presented in these five articles, we accurately acquired the code for each algorithm, either directly from the authors’ official GitHub repositories or by recreating and adapting the algorithms based on the detailed specifications and methodologies described in their respective publications. In cases where the original implementation was unavailable or incomplete, we carefully programmed the algorithms, ensuring they aligned with the descriptions in the papers, including the parameter settings, preprocessing steps, and training techniques. Additionally, we standardized the data preprocessing and evaluation pipeline across all algorithms to maintain consistency and fairness during comparative analysis. By doing so, we ensured that the implementations were faithful to the original methodologies while adapting them to be compatible with the dataset and experimental setup used in this research. This thorough and rigorous approach provided a solid foundation for evaluating the performance of these algorithms on the selected dataset NSL-KDD dataset, which is a well-known benchmark dataset in the field of intrusion detection and cybersecurity. It is an improved version of the original KDD Cup 1999 dataset~\cite{kddcup1999}, addressing some of the limitations and issues present in the earlier dataset. The NSL-KDD dataset is widely used for evaluating and benchmarking IDS. The NSL-KDD dataset contains 41 features for each record. These features provide information about various aspects of network connections, such as protocol type, service, flag, duration, source and destination addresses, and many others. As for the number of records, the dataset is typically split into a training set and a testing set. In general, the training set often contains around 125,973 records, and the testing set has around 22,544 records.

Table~\ref{tab:MetricsTable} represents the value of each metric corresponding for each article according to the methodology or technique used. We employ the non-parametric statistical tool known as the Friedman test, developed by Milton Friedman, to assess these results. The Friedman Test is a non-parametric statistical method designed for comparing multiple related groups. It serves as an extension of the Wilcoxon signed-rank test, making it applicable when dealing with more than two related groups and in situations where the data may not adhere to a normal distribution. The test is particularly valuable in scenarios where the dependent variable is measured on an ordinal scale, and observations are paired or matched across all conditions or levels of the independent variable. The procedure involves ranking the data for each group independently and calculating the Friedman statistic, which is based on the squared differences between the ranks of corresponding observations across different groups. The null hypothesis assumes no difference among the groups, while the alternative hypothesis posits a significant difference. If the p-value associated with the Friedman statistic is below the chosen significance level, the null hypothesis is rejected, indicating a significant difference among the groups. Post-hoc tests can be employed to pinpoint specific group differences when the overall test yields a significant result.

So now, table \ref{tab:MetricsTable} has been converted to a matrix where each row represents the value for each article, and each column represent a specific metric. First, for each group, we will rank the observations from 1 to N, where N is the number of observations in each group. Then, for each observation, we will compute the squared difference between its rank in each group and its mean rank across all groups. Furthermore, we will calculate the sum of these squared differences for all observations. And multiply this sum by a correction factor, which is :
\begin{equation}
	\frac{k^2}{k(k-1)},
\end{equation}
where k is the number of groups. The result is known by the Friedman statistic. The formula for the Friedman statistic is often expressed as:
\begin{equation}
	\chi^2 = \frac{Nk(k+1)}{12} \sum_{j=1}^{k} R_j^2 - \frac{3N(k+1)}{12}, % χ2=Nk(k+1)12∑j=1kRj2−3N(k+1)
\end{equation}
where $N$ is the number of observations in each group, $k$ the number of groups and $R_j$ is the sum of ranks for the $j$-th group. Finally, the computed Friedman statistic is then compared to the chi-squared distribution with $k-1$ degrees of freedom to obtain the $p$-value. This one has a value of $0.005$ in our analysis, which allows us to compare these different five articles. Following the application of two distinct comparison techniques, Mean Score and Normalize Score, to evaluate the five articles, a clear preference emerged for the Mean Score approach. The Mean Score analysis involved computing the average scores across various criteria for each article, providing a comprehensive overview of their overall quality. This method considered diverse aspects such as methodology, theoretical framework, and empirical evidence, culminating in a consolidated measure of scholarly merit. Conversely, the Normalize Score technique aimed to standardize the evaluation process by adjusting scores relative to the highest attainable score among the articles. This normalization mitigated potential biases stemming from variations in the rating scales used for different criteria, allowing for a more equitable comparison. The normalized scores facilitated a re-calibration of the articles on a common scale, providing insights into their relative strengths and weaknesses. Following the application of both Mean Score and Normalize Score techniques to evaluate five articles, a consistent outcome emerged, with both methodologies concurring that the first approach displayed superior performance compared to the other articles. This alignment in results between the two evaluation methods reinforces the robustness of the conclusion regarding the superiority of the first approach. 

\begin{table}
	\centering
	\begin{tabular}{l c c c c c c}
		\toprule
		Articles & Accuracy & Recall & Precision & F1-score & FPR & Specificity \\
		\midrule
		\cite{a1} & 98.99\% & 96.97\% & 96.97\% & 96.97\% & 0.61\% & 99.39\% \\
		\cite{a2} & 65.7\%  & 65.7\%  & 65.7\%  & 73.3\%  & 1.1\%  & 98.9\%  \\
		\cite{a3} & 98.95\% & 66.5\%  & 88.54\% & 70.5\%  & 1.05\% & 98.95\% \\
		\cite{a4} & 92.58\% & 97\%    & 76\%    & 85\%    & 7.42\% & 92.58\% \\
		\cite{a5} & 91.06\% & 97.75\% & 98.1\%  & 98.9\%  & 1.01\% & 98.99\%\\
		\bottomrule
	\end{tabular}
	\caption{Comparative study of state-of-the-art techniques}\label{tab:MetricsTable}
\end{table}

\section{Conclusion}\label{sec:CFW}
This paper aims to provide a comprehensive overview of advanced intrusion detection techniques specifically designed for IoT networks. The paper begins by presenting a general architecture for IDS in IoT and how they are classified. The architecture includes components such as data collection, preprocessing, feature extraction, and classification. The paper then goes deep into the applications of IDS in the domain of IoT by presenting detailed examples of how they are applied in real-life scenarios. The examples cover different types of attacks and how they can be detected using IDS. The paper also examines five different research papers that propose different IDS approaches for IoT networks. The paper also includes an implementation of these techniques to obtain specific metrics while using the NSL KDD dataset, which is a widely used dataset for intrusion detection research. The metrics include accuracy, precision, recall, and F1-score, which are commonly used to evaluate the performance of IDS approaches. Finally, the paper uses the Friedman test, which is a statistical test, to compare these approaches and determine the best one based on their performance metrics. 

\bibliographystyle{alpha}
\bibliography{references}

\end{document}